\documentclass[twocolumn,showpacs,preprintnumbers,amsmath,amssymb]{revtex4}

\usepackage{graphicx}
\usepackage{dcolumn}
\usepackage{bm}

\newcommand{\bv}[1]{{\boldsymbol #1}}
\newcommand{\bra}{\left<}
\newcommand{\ket}{\right>}
\newcommand{\chim}{\chi_{\rm m}}
\newcommand{\hc}{h_{\rm c}}
\newcommand{\hb}{h_{\rm b}}
\newcommand{\Rc}{R_{\rm c}}
\newcommand{\meq}{m_{\rm eq}}

\begin{document}


\title{Critical fluctuations of time-dependent magnetization in a random-field Ising model}

\author{Hiroki Ohta}
\email{hiroki@jiro.c.u-tokyo.ac.jp}
\author{Shin-ichi Sasa}%
 \email{sasa@jiro.c.u-tokyo.ac.jp}
\affiliation{%
Department of Pure and Applied Sciences, University of Tokyo, 3-8-1 Komaba, Meguro-ku 153-8902, Tokyo, Japan\\
}%


\begin{abstract}
Cooperative behaviors near the disorder-induced critical point in a random field Ising model are numerically investigated by analyzing time-dependent magnetization in ordering processes from a special initial condition.
 We find that the intensity of fluctuations of time-dependent magnetization,
 $\chi(t)$, attains a maximum value at a time $t=\tau$ 
in a normal phase and that $\chi(\tau)$  and $\tau$ exhibit divergences 
near the disorder-induced critical point. Furthermore, spin configurations
around the time $\tau$ are characterized by a length scale, which also 
exhibits a divergence near the critical point. We estimate the critical 
exponents that characterize these power-law divergences by using a 
finite-size scaling method.
\end{abstract}

\maketitle

\section{Introduction}

It has been known that earthquakes \cite{Sethna0}, acoustic emissions
in a deformed complex material \cite{Grasso}, 
and Barkhausen noise in random magnets\cite{Sethna1,Sethna2,Sethna3,Sabhapandit,Durin,Stanley}  exhibit distinctive power-law 
behaviors. All these systems possess 
some disorder and also they are 
driven by a slowly varying external field. As a simple model 
of such systems, a random field Ising model (RFIM) \cite{Young} 
under a slowly varying magnetic field has been investigated  
in order to elucidate the essential mechanism of the power-law behaviors.
By performing numerical experiments of the RFIM  under a slowly
 increasing magnetic field, it was found that the size 
distribution of avalanches, each of which represents a spin 
flipping in a connected region, becomes a power-law function 
at the critical strength of the disorder \cite{Sethna1}. 
Then, the critical magnetic field at which  the avalanche 
size becomes a system  size is called 
{\it disorder-induced critical point}.
Since this phenomenon 
occurs due to the existence of disorder, such power-law 
behaviors are called {\it disorder-induced critical phenomena}.

Despite the extensive studies for the critical phenomena,
fluctuations of magnetization, which might be the most
naive quantity characterizing the criticality, has never been 
investigated. Related to this issue, 
it has been known that the magnetization as a function of 
the magnetic field shows the almost 
discontinuous behavior at the disorder-induced critical 
point, although the precise determination of the 
transition type is still a controversy\cite{Sethna3,Vives2,Colaiori}. 
From this observation and considering the knowledge of 
conventional critical phenomena, static fluctuations 
of magnetization hardly exhibit singular behaviors 
near the disorder-induced critical point. These raise 
a naive question whether the disorder-induced critical 
point can be characterized in terms of fluctuations 
of time-dependent magnetization.

In this paper, we present a positive answer to this question. 
Our key idea is to notice the recent extensive 
studies for critical fluctuations near an ergodicity breaking 
transition in glassy systems \cite{Onuki,Bouchaud,Garrahan,Dauchot,Biroli,
 Miyazaki, Sasa}.
Here, we review these studies briefly. As an example, 
let us consider cooperative behaviors near the transition point in colloidal 
suspensions. In this system, near the transition point,
 there exist long-range spatial correlations 
among movable particles during a time interval 
$\tau$, which is chosen as a typical relaxation time. When
we introduce an appropriate quantity $Q(\bv{r},t)$ that indicates 
the occurrence of large particle displacement at the position 
$\bv{r}$ during the time interval $t$, a discontinuous jump
occurs in $Q(\bv{r},t\to \infty)$ accompanied with critical fluctuations
of $Q(\bv{r},\tau)$ at the ergodicity breaking transition. It should be noted
 that $Q(r,t)$ characterizes the dynamical event intrinsic to glassy systems. 

These results motivate us to study the cooperative behaviors near the 
disorder-induced critical point of the RFIM from the viewpoint 
of fluctuations of some dynamical events. 
In particular, we consider ordering processes of the 
magnetization from a special initial condition. We have numerically 
found that fluctuations of the time-dependent magnetization 
exhibit a critically divergent behavior near the disorder-induced 
critical point. 

This paper is organized as follows.
In Sec. \ref{pre}, we introduce the random-field Ising model and 
demonstrate its basic  behaviors numerically.  In Sec. \ref{result}, 
we address the main result of our study. Concretely, we 
present new critical exponents that characterize the phenomena
near the disorder-induced critical point. The final section is devoted
 to concluding remarks.

\section{Preliminaries}\label{pre}

Let us consider a three-dimensional cubic lattice 
$\Lambda \equiv \{\bv{i}=(x,y,z) | 1 \le x, y, z\le L \} $. 
A spin variable $\sigma_{\bv{i}} \in \{ -1,1 \} $ 
is defined at each site $\bv{i}\in \Lambda$. 
We study the RFIM described by the Hamiltonian
\begin{align}
H=-J\sum_{\bra \bv{i},\bv{j} \ket}
\sigma_{\bv{i}}\sigma_{\bv{j}}-\sum_{\bv{i} \in \Lambda}(h_{\bv{i}}+h)
\sigma_{\bv{i}},
\end{align}
where $\bra \bv{i},\bv{j} \ket$ represents a nearest-neighbor pair
of sites, $h$ is a constant external field and the random field 
$h_{\bv{i}}$ obeys a Gaussian distribution 
\begin{align}
P(h_{\bv{i}})=\dfrac{1}{\sqrt{2\pi R^2}}\exp(-\frac{h_{\bv{i}}^2}{2 R^2}).
\end{align}
The time evolution of the spin variables is described by 
the following rule. We first choose a site at random. If 
the spin flip on the site makes the system energy lower,  
the sign of the spin variable is changed; otherwise the 
spin flip is  rejected. At the next step, we choose 
a site at random again, and repeat the above-mentioned 
procedure. Here, a unit time is given by $L^3$ steps,
which is called a Monte Carlo step per site (MCS).
This time evolution rule is similar to traditional 
avalanche dynamics such as the one used in  Ref. \cite{Sethna3}, 
but our rule  may be closer to standard Glauber dynamics. 
Indeed, it corresponds to the Metropolis method with 
the zero temperature.  We should be careful to choose 
a time evolution rule particularly when we discuss finite 
temperature cases, but we do not enter a  difficult question 
which time evolution rule is more physical.

Here, we review the disorder-induced critical phenomena in
the RFIM. Let us consider the quasi static change of the external 
field from $h=-\infty$ to $h=\infty$.
In this operation with a given $R$, $\hc(R)$ is defined as the
special value of $h$  (if it exists) at which a spin flipping in 
a region over the whole system first occurs. Then, $\Rc$ is defined 
as the maximum value below which $\hc(R)$ exists. 
Their actual values were numerically 
determined as $(\Rc,\hc(\Rc)) \simeq(2.16,1.44)$ \cite{Sethna3}, which 
is called disorder-induced critical point. 
It was reported that power-law behaviors of the size distribution of 
avalanches were observed in the quasi static change of the 
magnetic field when $R=R_c$. Such power-law
behaviors are called the disorder-induced critical phenomena.

\begin{figure}
\includegraphics[width=7.5cm,clip]{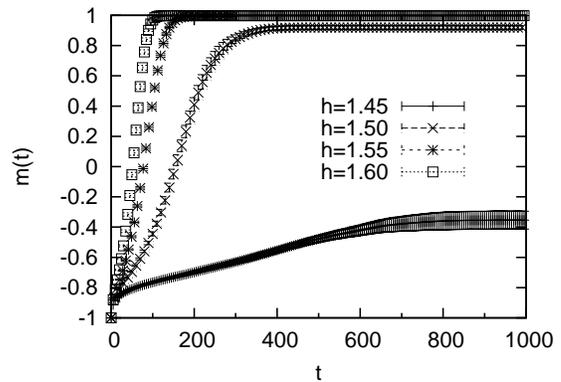}
\caption{Ordering process of magnetization, $m(t)$. 
$L = 40$ and $R=2.16$}
\label{magnet}
\end{figure}

In this paper, we study ordering processes of
 magnetization 
\begin{align}
\hat{m}(t)&\equiv\frac{1}{N}\sum_{\bv{i} \in \Lambda} \sigma_{\bv{i}}(t)
\end{align}
from the initial condition $\hat{m}(0)=-1$ in which all 
the spins are downward. We also focus on 
the case $T=0$, except for a brief discussion in the final section. 
As preliminary calculations, we measure 
\begin{align}
m(t)&\equiv \bra  \hat{m}(t) \ket 
\end{align}
for several values of $(R,h)$, 
where $\bra  A \ket$ represents the average of a 
physical quantity $A$ with respect to the stochastic
time evolution and quenched disorder.
In the argument given below, we consider at least 20 samples of time evolution 
for each set of $\{h_{\bv i}\}$
and at least 20 samples of $\{h_{\bv i}\}$ for calculating average values.

\begin{figure}
\includegraphics[width=7.5cm,clip]{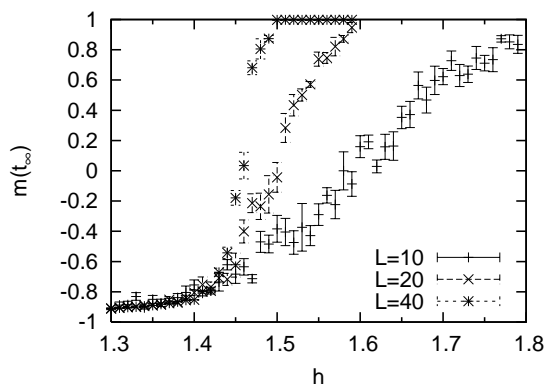}
\caption{ $m(t_{\infty})$ as a function of $h$
for systems with three different sizes.}
\label{order}
\end{figure}

Typical samples of $m(t)$ for the case $R=2.16$ 
are displayed in Fig. \ref{magnet}. 
It is observed that the magnetization $m(t)$ quickly approaches
the equilibrium value $\meq$, which is approximately 1, from -1 when $h$ is 
sufficiently large and that the ordering process becomes 
slower for the system with smaller $h$. Note that 
the equilibrium value $\meq$ is slightly less than $1$
because $h_i$ on some sites, whose values are largely negative, 
prevent the magnetization from approaching 1. Then, by decreasing $h$ further, 
we find that there is a value below which $m(t)$ does not 
reach the state with the equilibrium value $\meq$.

\begin{figure}
\includegraphics[width=7.5cm,clip]{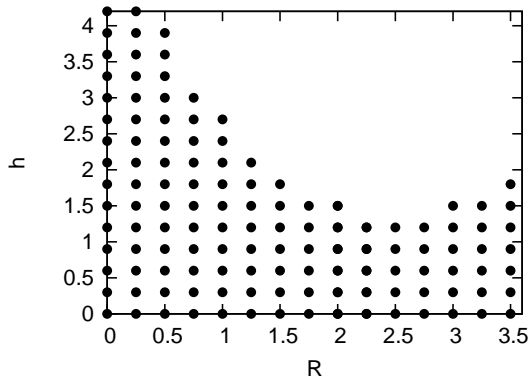}
\caption{Phase diagram. The circles 
represent frozen states. $L=40$.}
\label{phase}
\end{figure}

In order to quantify this transition, 
we consider the quantity $m(t_\infty)$, where $t_{\infty}$ 
is chosen as 1000 MCSs in our numerical experiments. We confirmed
that the results reported below did not depend on the choice of 
$t_\infty$ when $t_\infty \ge 1000$ MCSs. (The lowest value of $t_\infty$
depends on the system size.) On the basis of this, we define 
the {\it frozen phase} as the state with $m(t_{\infty}) < \meq$. 
Since a discontinuous transition from $\meq$ to $m(t_\infty) $ 
is expected to occur at a certain value of $h$ in the large size limit, 
we can replace the exact definition of the frozen phase with an 
operational one expressed as $m(t_\infty) < 1-\epsilon$, where $\epsilon=0.1$ 
for numerical simplicity. Indeed, as shown in Fig. \ref{order},
the transition becomes sharper when $L$ is larger. Therefore  we 
expect that our operational definition provides an accurate determination 
of the phases in the large size limit. 
Presently, we fix $L=40$ and test whether 
the system exhibits a frozen phase for several values of $(R,h)$.
The result is summarized in Fig. \ref{phase}, which 
suggests the presence of a transition curve $\hb(R,L,\epsilon)$ 
between the frozen and unfrozen phases. Note that the disorder-induced 
critical point $(\Rc,\hc)$ reported in Ref. \cite{Sethna3} appears to be
located on this curve. More precise correspondences with previous phase diagrams
reported in Refs. \cite{Sethna2,Muller,Vives} will be discussed elsewhere. 

\section{Result}\label{result}

Now, we focus on the behaviors near the disorder-induced 
critical point $(\Rc,\hc)$. That is, by fixing $R$ as $\Rc$, we 
investigate the system with several values of $h$ near $\hc$.  
Following our motivation, we are interested in studying fluctuations 
of $\hat m(t)$. It should be noted that  
our study is concerned with {\it fluctuations of relaxation events}.  
Although such types of fluctuations of dynamical events have
been studied extensively in an ergodicity breaking transition in glassy systems, to our knowledge, there have been no such arguments on statistical properties near the disorder-induced critical point. 

\begin{figure}
\includegraphics[width=7.5cm,clip]{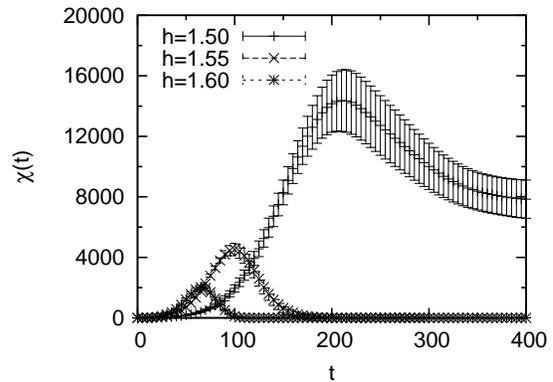}
\caption{$\chi(t)$ for three values of $h$. $R=2.16$.}
\label{chi}
\end{figure}

Since the simplest quantity characterizing the fluctuations of 
$\hat m(t)$ is given by 
\begin{align}
\chi (t) &\equiv N\left[\left< \hat{m}(t)^2 \right> - \bra \hat{m}(t) \ket^2\right],
\end{align}
we first demonstrate the graphs of $\chi(t)$ for a few values of $h$
in Fig. \ref{chi}. It is observed that  $\chi(t)$ has a peak at a time 
$\tau$ and that both $\chim=\chi(\tau)$ and 
$\tau$ increase  when $h$ approaches $\hc (\simeq 1.44)$.  
Based on this observation, we next attempt to extract divergent behaviors
of $\chim$ and $\tau$ by using a finite-size scaling analysis. 

\begin{figure}
\centering
 \includegraphics[width=7.5cm]{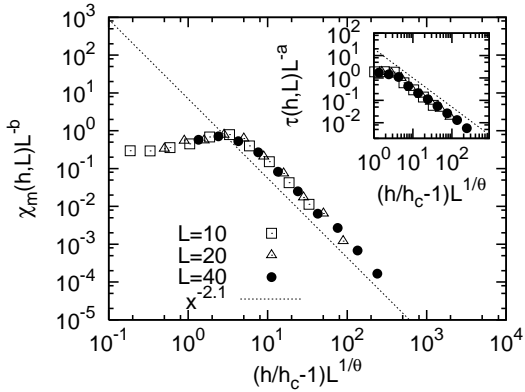}
\caption{Finite-size scaling for $\chi_m(h,L)$ and $\tau(h,L)$ (inset).
 The three graphs for systems with different sizes are collapsed into a single curve. The statistical error bars are within the size of symbols.
$\hc=1.44$ and $R=2.16$.}
\label{chiRc}
\end{figure}

Thus for the systems with $L=10$, $20$, and $40$, 
we measured $\tau(h,L)$ as a function of $(h-\hc)/\hc$. 
As shown in the inset of Fig. \ref{chiRc}, when we plot $\tau(h,L) 
L^{-a}$ as a function of $L^{1/\theta} (h-\hc)/\hc$, these three graphs 
do not depend on $L$, where $a$ and $\theta$ are fitting parameters whose 
values ($\theta \simeq 0.7$ and $a \simeq 1.7$) are 
determined in such a manner that the three graphs are collapsed into a single curve as exactly as possible. 
Based on this result, we conjecture a scaling form
\begin{align}
\tau(h,L) &= L^aF_{\tau}\left(\left|\dfrac{h-\hc}{\hc} \right|
L^{\frac{1}{\theta}} \right),
\end{align}
by using the scaling function $F_\tau$. Considering the
asymptotic law $F_\tau (z) \simeq z^{-\zeta}$ (with $\zeta
\simeq 1.3)$ in the regime $z \gg 1$, we expect the following
critical behavior in the large size limit:
\begin{equation}
\tau \simeq (h-\hc)^{-\zeta}.
\label{tau:s}
\end{equation}

In a manner similar to that in the analysis of $\tau(h,L)$, we assume
a form of the finite-size scaling as follows:
\begin{align}
\chi_m(h,L) &= L^bF_{\chi}\left(\left|\dfrac{h-\hc}{\hc}
\right|L^{\frac{1}{\theta}}\right). 
\end{align}
Indeed, from Fig. \ref{chiRc}, we determine $b$ 
and the scaling function $F_\chi$, where $b \simeq 3.0$ 
and we find the asymptotic relation $F_\chi (z) \simeq z^{-\gamma}$ 
(with $\gamma \simeq 2.1$) in the regime $z \gg 1$. 
We thus obtain
\begin{equation}
\chim \simeq (h-\hc)^{-\gamma}.
\label{chi:s}
\end{equation}

\begin{figure}
\centering
\includegraphics[width=7.5cm,clip]{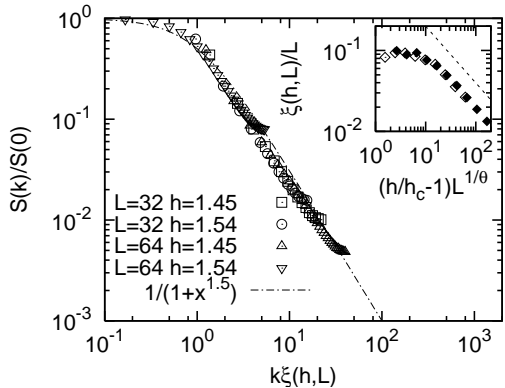}
\caption{$S(\bv{k})/S(\bv{0})$ as a function of $k\xi(h,L)$ Inset: Finite-size scaling for $\xi(h,L)$. Filled symbols represent those of the system for L=64 and the others for L=32. $R=2.16$.}
  \label{L}
\end{figure}

These divergent behaviors observed near the disorder-induced critical point
lead us to expect that some spins flip cooperatively around the 
time $\tau$. In order to describe the nature of the cooperative  
phenomena,  we attempt to define a spatial correlation length that characterizes 
it, as carried out in studies on traditional critical phenomena.  
Here, it should be noted that the magnetization grows around the time $\tau$.
In this paper, for numerical simplicity, we focus on the spin 
configurations at the time $t_0$ such that $\hat{m}(t_0)=0$ for each sample
because the magnetization is 
expected to grow at this time. We then consider the 
spatial pattern indicating whether the spin on each site has already 
flipped by the time $t_0$. Thus, we measure 
\begin{align}
\rho(\bv{j}) & = \delta(\sigma_{\bv{j}}(t_0),1),
\end{align}
where $\delta(m,n)$ represents  Kronecker's delta.
Calculating its Fourier transform 
\begin{align}
\tilde \rho(\bv{k}) & = \sum_{\bv{j} \in \Lambda}
\rho( \bv{j} ) \exp( i \bv{k}\cdot\bv{j} ),
\end{align}
we define the structure function 
\begin{align}
S(\bv{k}) & \equiv \frac{1}{N}\left< |\tilde{\rho}(\bv{k})|^2 
\right>, 
\end{align}
where we set $\bv{k}=(k,0,0)$ in the argument given below.
For $S(\bv{k}) $ obtained for several values of $(h,L)$,
we find that the fitting 
\begin{align}
S(\bv k) = \frac{S(\bv{0})}{1+(\xi k)^{n}}
\end{align} 
is obtained well with $n=3/2$, as shown in Fig. \ref{L}. 
This is called the Ornstein-Zernike form if 
$n=2$ \cite{Onuki}. Then,  $\xi$ represents the correlation length 
characterizing the spatial pattern $\rho(\bv{i})$.  

Now, in the same manner as those for $\tau$ and $\chim$,
we perform a finite-size scaling analysis assuming
the form 
\begin{align}
\xi(h,L) &= LF_{\xi}
\left(\left|\dfrac{h-\hc}{\hc} \right|L^{\frac{1}{\theta}} \right).
\end{align}
The inset of Fig. \ref{L} illustrates that this assumption
is reasonable and that $F_{\xi}(z)$ obeys the asymptotic relation 
$F_{\xi}(z) \simeq z^{-\nu}$ for large  $z$ ($\nu \simeq 0.7$). 
On the basis of this, we obtain the critical behavior of the 
dynamical correlation length:
\begin{align}
\xi \simeq |h-\hc|^{-\nu}.
\label{xi:s}
\end{align}
Note that the obtained value $\nu \simeq 0.7$ is close to the 
 value $\theta \simeq 0.7$, where $\theta$ is the exponent that
characterizes the length scale appearing in the finite-size 
scaling method. This coincidence implies that the manner of the divergence
 for length scales is characterized by a single exponent.
It should be noted that the exponents characterizing the 
length scale appearing in finite-size scaling analysis
 for statistical quantities related to avalanches were obtained 
in Refs. \cite{Dahmen,Vives2}, where the values of the 
exponent are close to that of $\theta$ in our study. The relation 
among these results  will be studied in the future.

One may be afraid that another time evolution rule provides
a different result, in particular with regard to the exponent 
$\zeta$. Until now, we do not understand its dependency, but we 
conjecture that the value of $\nu$ is not so influenced  
by the choice of the time evolution rule, because of the 
consistency with the previous studies. These will be also 
studied in the future.
\section{Concluding remarks}

\begin{figure}
\includegraphics[width=5cm,clip]{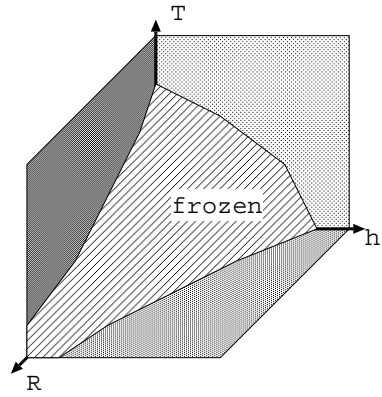}
  \caption{Schematic phase diagram in the (R,H,T) space. }
  \label{sketch}
\end{figure} 


We have presented the power-law divergences (\ref{tau:s}), 
(\ref{chi:s}), and (\ref{xi:s}) near the disorder-induced critical 
point in the RFIM. We have determined the exponents by using the finite-size
 scaling method as $\gamma \simeq 2.1$, $\zeta \simeq 1.3$ and 
$\nu \simeq 0.7$. These values are regarded as preliminary values
and more precise values will be determined by using systems with
considerably larger sizes. At present, the accuracy of values is not of primary interest,
 but the existence of divergent fluctuations of time-dependent magnetization 
is rather important. Indeed, based on our results, 
we conclude that the phenomena near 
the disorder-induced critical point can be captured from  the 
viewpoint of fluctuations of dynamical events. In this sense, 
the phenomena under consideration have common features  with 
cooperative behavior in glassy and jamming systems. It is an 
important future subject whether the  values of critical exponents 
observed near the disorder-induced critical point are related 
to those in glassy and jamming systems.


Last, we introduce a few examples of 
studies motivated by this conclusion. The first 
example is the phase diagram in the $(R,H,T)$ space. Since the 
precise definition of the frozen phase appears to be complicated
in the finite temperature case, we just plotted the region $m(t_{\infty}) < 0$ as a tentative 
frozen phase, using a Metropolis method.
Figure \ref{sketch} shows a schematic diagram of this phase. 
From this figure, one may recall the phase diagram of 
the jamming transition in granular systems (see Fig. 1 
in Ref. \cite{Nagel} or Fig. 4 in Ref \cite{Weitz}). 
 We also expect that a similar 
type of phase diagram can be obtained when we employ other 
time evolution rules such as a heat bath method.
Thus this resemblance motivates us to 
study the common aspects between granular systems and the 
present system.

The second example is related to a theoretical framework.
In addition to extensive analysis on the power-law distribution of avalanches 
\cite{Sethna2,Sabhapandit,Stanley}, it was conjectured that the critical 
behaviors of avalanches 
in some spin models with disorder are related to metastable states
\cite{Horbach,Zimanyi,Tarjus}. It is interesting to investigate 
$\chi(t)$ in such systems. 
Furthermore, we are interested 
in conducting a theoretical analysis of our numerical results. 
Since $\chi(t)$ proposed in this paper has never been studied 
in the RFIM, such a theoretical analysis would shed light on a
new aspect of the RFIM.

\begin{acknowledgments}
The authors thank K. Hukushima for useful comments on this work
 and E. Vives for telling us about Ref. \cite{Vives2} with many useful 
comments. This work was supported by a grant from 
the Ministry of Education, Science, Sports and Culture of Japan (Grant No. 19540394). 
\end{acknowledgments}

\end{document}